# Hydrogenation Dynamics of Biphenylene Carbon (Graphenylene) Membranes

Vinicius Splugues[1], Pedro Alves da Silva Autreto[1,2], Douglas S. Galvao[1]
[1]Instituto de Física "Gleb Wataghin", Universidade Estadual de Campinas, Campinas - SP, 13083-970, Brazil
[2]Universidade Federal do ABC, Santo André-SP, 09210-580, Brazil

**ABSTRACT**

The advent of graphene created a revolution in materials science. Because of this there is a renewed interest in other carbon-based structures. Graphene is the ultimate (just one atom thick) membrane. It has been proposed that graphene can work as impermeable membrane to standard gases, such argon and helium. Graphene-like porous membranes, but presenting larger porosity and potential selectivity would have many technological applications. Biphenylene carbon (BPC), sometimes called graphenylene, is one of these structures. BPC is a porous two-dimensional (planar) allotrope carbon, with its pores resembling typical sieve cavities and/or some kind of zeolites. In this work, we have investigated the hydrogenation dynamics of BPC membranes under different conditions (hydrogenation plasma density, temperature, etc.). We have carried out an extensive study through fully atomistic molecular dynamics (MD) simulations using the reactive force field ReaxFF, as implemented in the well-known Large-scale Atomic/Molecular Massively Parallel Simulator (LAMMPS) code. Our results show that the BPC hydrogenation processes exhibit very complex patterns and the formation of correlated domains (hydrogenated islands) observed in the case of graphene hydrogenation was also observed here. MD results also show that under hydrogenation BPC structure undergoes a change in its topology, the pores undergoing structural transformations and extensive hydrogenation can produce significant structural damages, with the formation of large defective areas and large structural holes, leading to structural collapse.

**INTRODUCTION**

Topologically, graphene can be considered as a two-dimensional, one-atom thick membrane formed by an array of hexagonal $sp^2$ bonded carbon atoms (Figure 1A) [1-4]. It has been theoretically investigated since late 1940s as a model to describe some properties of graphite. After Novoselov and Geim, using a "scotch tape" method, experimentally obtained a single layer of graphene from highly oriented pyrolytic graphite (HOPG) [4], an extraordinary number of theoretical and experimental works has been published on this material. Although graphene presents several remarkable properties, there are some difficulties to be overcome before real graphene-based nanoelectronics can become a reality. These difficulties are mainly related to its zero bandgap band structure (semi-metal), which precludes its direct use for some devices, such as digital transistors and diodes [5].

Among the many possible applications of graphene [6,7], it has been proposed its use as a selective membrane for water filtration or some gases. However, graphene in its defectless form (pristine) has been shown to be impermeable even to the smallest species, such as noble gases (argon and helium [1]). Due in part to this, there is a renewed interest in other carbon-based, graphene-like structures but exhibiting larger porosity. Graphynes [8,9] (already synthesized) and biphenylene carbon (BPC), also known as graphenylene (Figure 1b) [8,10], are good candidates satisfying these requirements, because they present larger porosity than graphene and potential selectivity. BPC is a two-dimensional structure with an interesting topology, with pores that resemble some types of zeolites [2,10]. An equivalent BPC inorganic structure also exists [11].

To be used as a sieve or selective membrane, the interaction between the solvent and the membranes should be only van der Waals, once reactions (covalent bonds) could completely destroy the membrane topology and its filtration capabilities. For a better understanding of these aspects, in this work, we investigated the BPC hydrogenation dynamics [12], under different conditions.

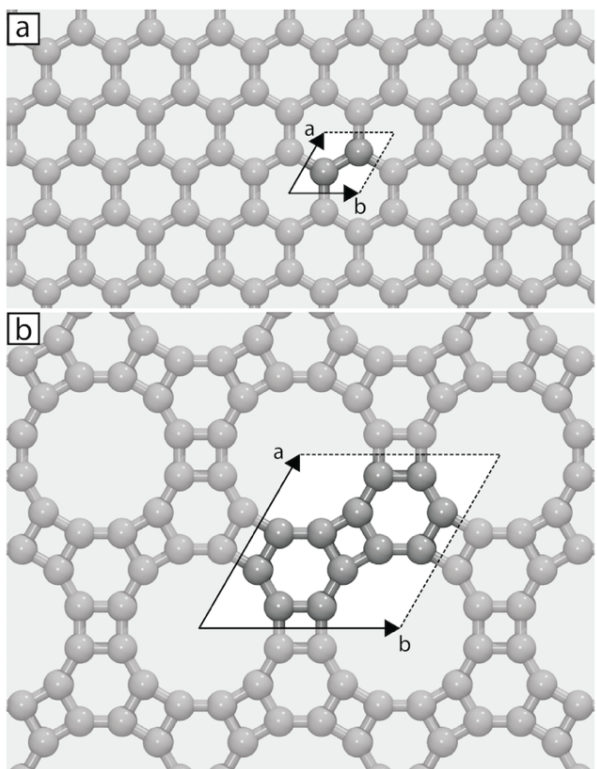

**Figure 1**. (a) Graphene and; (b) biphenylene carbon (BPC) (b) structures. Their structural units cells are highlighted in white color.

**THEORY**

To study the BPC hydrogenation processes, we used fully atomistic reactive molecular dynamics (MD) simulations. MD is a technique where we basically solve Newton's equations for the investigated system and analyze its time evolution in terms of forces and energies [13-14].

To simulate the hydrogenation process there is a need to use reliable force fields to address the formation and/or breaking of chemical bonds. This can be done using the reactive force field called ReaxFF [15-16], already successfully tested in other hydrogenation processes

[3,12]. The simulations were carried out using the public-domain computational code known as the Large-scale Atomic / Molecular Massively Parallel Simulator (LAMMPS) [17].

The systems considered in our simulations were composed of BPC membranes (initially with dimensions about 120 Å x 120 Å, containing approximately 5000 carbon atoms) and immersed into an atmosphere of hydrogen atoms (Figure 2). A constant volume simulation box (NVT ensemble) was used. We considered different hydrogen atmosphere density values and different temperatures. The simulation average time was ~100 ps, using timesteps of 0.1 fs (corresponding to ~106 MD steps).

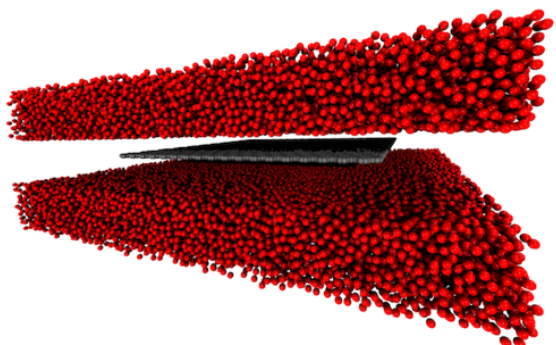

**Figure 2**. Typical system considered in the MD simulations. Red (light) and gray (darker) colors indicated hydrogen and carbon atoms, respectively.

In order to gain further insights on the existence of favored hydrogenation sites, we have also computed 3D potential energy maps. These maps were generated considering potential energy changes experienced by a probe hydrogen atom located at 1.5 Å above the BPC basal plane. Negative energy values mean energetically favorable regions for hydrogen bond formation.

**DISCUSSION**

One effective way to visualize the relative importance of the sites to be hydrogenated is from the above-mentioned 3D energy potential maps. These maps provide information about which sites can be preferentially attacked. In Figure 3 we present these maps for a BPC membrane and its hydrogenated form.

As we can see from these maps, initially there are not significant differences among the possible sites in terms of preferential hydrogen bonding (Figure 3a). However, a single hydrogen atom attached to BPC membrane (Figure 3b) is enough to significantly alter the maps. This means that once a hydrogen bond is formed, there is an increased probability of neighbor carbon atoms to be also hydrogenated, favored a fast hydrogenation dynamics.

In Figure 4 we present the number of atoms bonded for three different temperatures 300K, 450K and 600K as a function of time simulation. As we can see from this Figure, the hydrogenation processes are highly dependent on temperature. As we increase the temperature the number of hydrogen atoms incorporated into the structure considerably increases. Our results also show that the hydrogenation occurs with the formation of correlated domains (islands of hydrogenated carbons) very similar to what was reported to graphane (hydrogenated graphene) [3], but quite distinct from the case of fluorinated graphene [12]. This could be a consequence of the porous BPC structure, which allows larger out-of-plane deformations, in comparison to graphene. These large out-of-plane deformations results in a fast hydrogenation dynamics, with the simultaneous formation of many partially hydrogenated regions, which lead to the formation of correlated domains, as in the case of graphene [3].

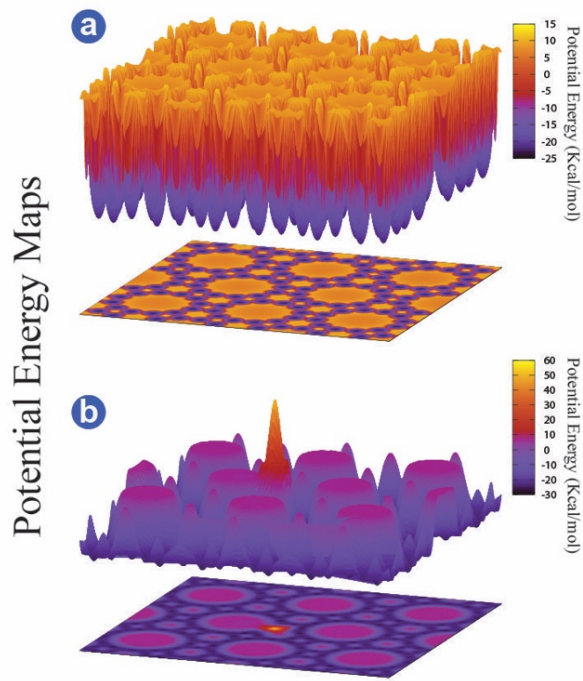

**Figure 3**. Potential energy maps. (a) Pristine BPC; (b) Hydrogenated BPC. See text for discussions.

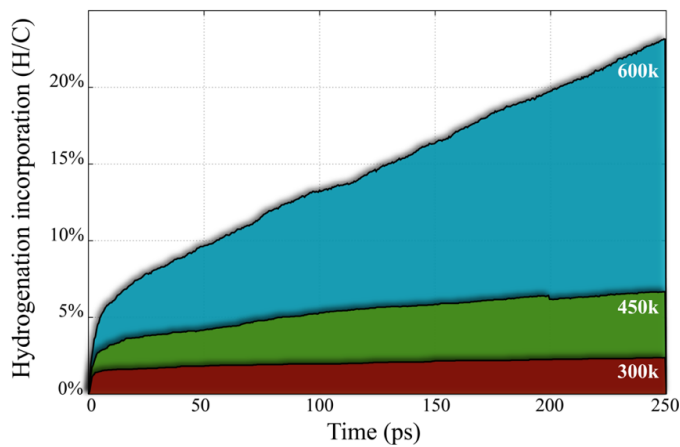

**Figure 4**. Rate of the hydrogen incorporation at different temperatures, as a function of time simulation.

More pronounced extensive hydrogenation occurred for higher temperatures and can cause serious structural damages, as we can see in the case of hydrogenation at 800 K (Figure 5). For this temperature, the structure collapsed, i. e., the structural integrity of the membrane is compromised through the formation of major structural defects and "holes". This is similar to what was previously observed for graphene [3, 12, 18].

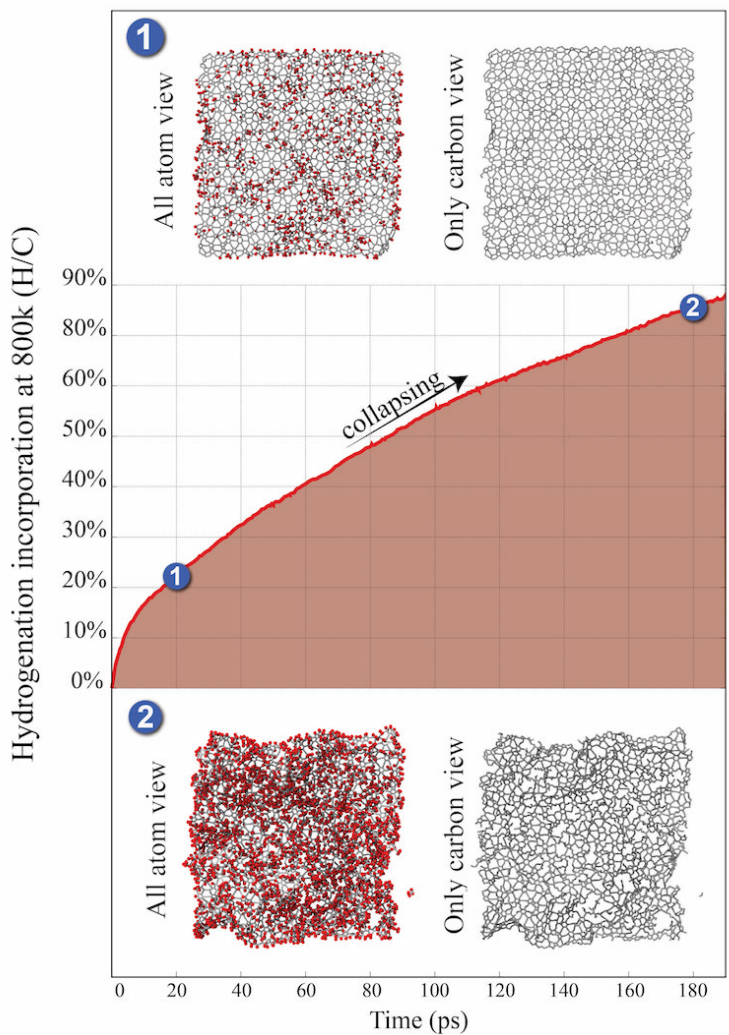

**Figure 5**. Rate of hydrogen incorporation at 800K, as a function of time simulation. Representative MD snapshots for the indicated time values (1 and 2, respectively) are also displayed.

**CONCLUSIONS**

We performed reactive molecular dynamics simulations of the hydrogenation processes of biphenylene carbon (BPC), also known as graphenylene. Our results showed that extensive hydrogenation at high temperatures can produce significant structural damages, with the formation of large defective areas and large structural holes, which could result in structural collapse. Our results also show that the hydrogenation occurs with the formation of correlated domains (islands of hydrogenated carbons) very similar to the case of graphane (hydrogenated graphene) [3]. This could be a consequence the larger BPC porosity (in relation to graphene), which allows large out-of-plane deformations. These large out-of-plane deformations results in a fast hydrogenation dynamics, as the chemical reactivity is proportional to local curvature [3, 12, 18]. This results in the simultaneous formation of many partially hydrogenated regions, which lead to the formation of correlated domains, as in the case of graphene [3].

## ACKNOWLEDGMENTS

Work supported in part by the Brazilian Agencies CAPES and CNPq. Computational and financial support from the Center for Computational Engineering and Sciences at Unicamp through the FAPESP/CEPID Grant No. 2013/08293-7 is also acknowledged.

## REFERENCES

[1] J. S. Bunch, S. S. Verbridge, J. S. Alden, A. M. van der Zande, J. M. Parpia, H. G. Craighead, and P. L. McEuen, *Nano Lett.* **8**, 2458 (2008).
[2] G. Brunetto, P. A. S. Autreto, L. D. Machado, B. I. Santos, R. P. B. Dos Santos e D. S. Galvao, *J. Phys. Chem. C* **116**, 12810 (2012).
[3] M. Z. S. Flores, P. A. S. Autreto, S. B. Legoas e D. S. Galvao, *Nanotechnology* **20**, 465704 (2009).
[4] K. S. Novoselov, A. K. Geim, S. V. Morozov, D. Jiang, Y. Zhang, S. V. Dubonos, I. V. Grigorieva, and A. A. Firsov, *Science* **306**, 666 (2004).
[5] F. Withers, M. Dubois, and A. K. Savchenko, *Phys. Rev. B* **82**, 73403 (2010).
[6] I. W. Franck, D. M. Tanenbaum, A. M. van der Zende, and P. L. McEuen, *J. Vaccum Sci. & Technol. B* **25**, 2558 (2007).
[7] R. Faccio, P. A. Denis, H. Pardo, C. Goyenola, and A. W. Mombru, *J. Phys. Cond. Mat.* **21**, 285304 (2009).
[8] R. H. Baughman, H. Eckhardt e M. Kertesz, *J. Chem. Phys.* **87**, 6687 (1987).
[9] V. R. Coluci, S. F. Braga, S. B. Legoas, D. S. Galvao e R. H. Baughman, *Phys. Rev. B* **68**, 035430 (2004).
[10] G. Brunetto and D. S. Galvao, *MRS Proc.* **1658**, mrsf13-1658-rr07-20 (2014).
[11] E. Perim, R. Paupitz, P. A. S. Autreto, and D. S. Galvao, *J. Phys. Chem.* **C118**, 23670 (2014).
[12] R. Paupitz, S. B. Legoas, S. G. Srinivasan, A. C. T. van Duin, and D. S. Galvao, *Nanotechnology* **24**, 035706 (2013).
[13] C. J. Casewit, K. S. Colwell e A.K. Rappé, *J. Am. Chem. Soc.* **114**, 10035 (1992).
[14] L. Kale, R. Skeel, M. Bhandarkar, R. Brunner, A. Gursoy, N. Krawetz, J. Philips, A. Shinozaki, K. Varadarajan e Klaus Shulten, *J. Comp. Phys.* **151**, 283 (1999).
[15] A. C. T. van Duin, S. Dasgupta, F. Lorant e W. A. Goddard III, *J. Phys. Chem. A*. **105**, 9396 (2001).
[16] S. G. Srinivasan, A. C. T. van Duin and P. Ganesh, *J. Phys. Chem. A*. **119**, 571 (2015).
[17] S. Plimpton, *J. Comp. Phys.* **117**, 1 (1995).
[18] G. M. Psofogiannakis and G. E. Froudakis, *J. of Phys. Chem. C*, **116(36)**:19211 (2012).